\documentclass[twocolumn,pra,twocolumn,superscriptaddress]{revtex4}
\usepackage{color}
\usepackage{amsmath}
\usepackage{amssymb}
\usepackage{graphicx}
\usepackage{tikz}
\usepackage{multirow}
\usepackage{float}
\usepackage{bbm}
\usepackage{dsfont}
\usepackage{cases}
\usepackage{mathtools}

\usepackage{epsfig}
\usepackage{verbatim}
\usepackage{array}
\usepackage{setspace}

\usepackage{scalerel}
\usepackage{stackengine,wasysym}
\usepackage{empheq, nccmath}

\usepackage[T1]{fontenc}

\newcommand\blue[1]{{\color{black}#1}}

\usepackage[unicode=true,
bookmarks=true,bookmarksnumbered=false,bookmarksopen=false,
breaklinks=false,pdfborder={0 0 1},backref=false,colorlinks=true]
{hyperref}
\hypersetup{
	linkcolor=magenta, urlcolor=blue, citecolor=blue, pdfstartview={FitH}, hyperfootnotes=false, unicode=true}

\makeatletter
\@ifundefined{textcolor}{}
{%
	\definecolor{BLACK}{gray}{0}
	\definecolor{WHITE}{gray}{1}
	\definecolor{RED}{rgb}{1,0,0}
	\definecolor{GREEN}{rgb}{0,1,0}
	\definecolor{BLUE}{rgb}{0,0,1}
	\definecolor{CYAN}{cmyk}{1,0,0,0}
	\definecolor{MAGENTA}{cmyk}{0,1,0,0}
	\definecolor{YELLOW}{cmyk}{0,0,1,0}
}

\begin{document}

\preprint{APS/123-QED}

\title{Free-fermion models and the two-dimensional Ising models under the zero field and imaginary field $i(\pi/2){k_B}T$}

\author{De-Zhang Li}
\address {Quantum Science Center of Guangdong-Hong Kong-Macao Greater Bay Area, Shenzhen 518045, China}
\author{Xin Wang}
\thanks{Corresponding author: x.wang@cityu.edu.hk}
\address {Department of Physics, City University of Hong Kong, Hong Kong SAR, China}
\author{Xiao-Bao Yang}
\thanks{Corresponding author: scxbyang@scut.edu.cn}
\address {Department of Physics, South China University of Technology, Guangzhou 510640, China}

\begin{abstract}
Ising model is famous in condensed matter and statistical physics. In this work we present a free-fermion formulation of the two-dimensional classical Ising models on the honeycomb, triangular and Kagomé lattices. Each Ising model is studied in the cases of a zero field and of an imaginary field $i( {{\pi  \mathord{\left/{\vphantom {\pi  2}} \right. \kern-\nulldelimiterspace} 2}} ){k_B}T$. We employ the decorated lattice technique, star-triangle transformation and weak-graph expansion method to exactly map each Ising model in both cases into an eight-vertex model on the square lattice. The resulting vertex weights are shown to satisfy the free-fermion condition. In the zero field case, each Ising model is an even free-fermion model. In the case of the imaginary field, the Ising model on the honeycomb lattice is an even free-fermion model while the models on the triangular and Kagomé lattices are odd free-fermion models. We obtain the exact solution of the Kagomé lattice Ising model under the imaginary field $i( {{\pi  \mathord{\left/{\vphantom {\pi  2}} \right. \kern-\nulldelimiterspace} 2}} ){k_B}T$, a result not previously reported in the literature. We also show that the frustrated Ising models on the triangular and Kagomé lattices in the imaginary field still exhibit a non-zero residual entropy.
\end{abstract}

\keywords{Ising model, sixteen-vertex model, free-fermion condition, exact solution}
\maketitle

\section{\label{sec:level1}Introduction}
Ising model plays a crucial role in the research of condensed matter and statistical physics of lattice systems, especially in understanding the phase transition via statistical mechanics. It has a long history since 1920s \cite{RN312}. The one-dimensional Ising model was first solved by Ising himself \cite{RN75}. Kramers and Wannier introduced the transfer matrix method to rederive the exact solution for the one-dimensional case and to study the two-dimensional case \cite{RN236, RN237}. 
Onsager was the first to present the exact solution of the two-dimensional Ising model on the square lattice without a magnetic field, using the transfer matrix method \cite{RN72}. This famous result has since been obtained through various approaches \cite{RN73, RN267, RN74, RN76, RN207, RN265, RN269, RN218, RN209}, and has made a great impact on exactly solvable lattice models \cite{RN239, RN49, RN205, RN152, RN206}.  Lee and Yang proposed the solution for the case in an imaginary field $i( {{\pi  \mathord{\left/{\vphantom {\pi  2}} \right. \kern-\nulldelimiterspace} 2}} ){k_B}T$ in their work of Lee-Yang zeros \cite{RN57}. This solution has also been rederived using a variety of different methods \cite{RN67, RN71, RN68, RN69, RN70, RN66, RN274}.
Since then, the mathematically exact results for Ising models have attracted considerable attention, demonstrating their elegance and significance. 
Two-dimensional Ising models on the honeycomb lattice \cite{RN122, RN123, RN221, RN224}, the triangular lattice \cite{RN81, RN220, RN299, RN223, RN264}, the Kagomé lattice \cite{RN121, RN82}, and the checkerboard lattice \cite{RN58, RN51} were also exactly solved. Although the Ising model in a non-zero field is generally a well-known unsolved problem, the case of an imaginary field $i( {{\pi  \mathord{\left/{\vphantom {\pi  2}} \right. \kern-\nulldelimiterspace} 2}} ){k_B}T$ introduced by Lee and Yang \cite{RN57}, is unique and holds significant interest.

In this work we focus on the exact solutions of anisotropic Ising models on typical two-dimensional lattices, both in the zero field and in the imaginary field. The Hamiltonian of \textit{N} spins $\left\{ {{s_i} =  \pm 1,{\rm{ }}i = 1, \cdots ,N} \right\}$  reads
\begin{equation}
H = \sum\limits_{\left\langle {ij} \right\rangle } {J_{ij}{s_i}{s_j}}  - {H_{{\rm{ex}}}}\sum\limits_{i = 1}^N {{s_i}}~,  \label{eq1}
\end{equation}
where the first sum $\sum_{\left\langle {ij} \right\rangle } {}$ is over all nearest-neighbours, \blue{$J_{ij}$ denotes the directional anisotropic interaction} and the field ${H_{{\rm{ex}}}}$ can be 0 or $i( {{\pi  \mathord{\left/{\vphantom {\pi  2}} \right. \kern-\nulldelimiterspace} 2}} ){k_B}T$. The approach we employ is the mapping into the free-fermion model \cite{RN65, RN59, RN129, RN262, RN288}, which was proposed in the vertex model problem. 

\begin{figure*}
\includegraphics{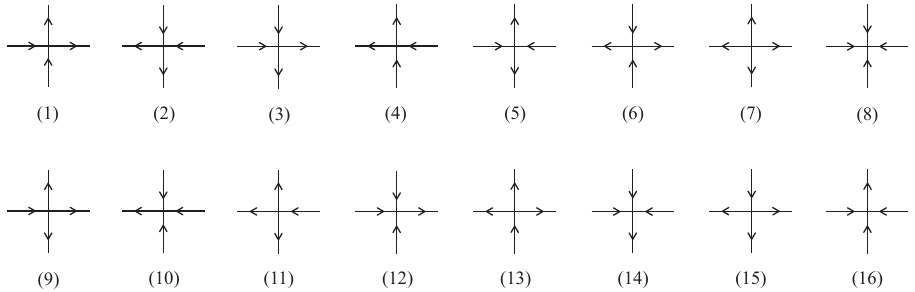}
\caption{\label{fig1} The vertex configurations of the sixteen-vertex model. The even subcase consists of vertices (1)--(8), while the odd subcase consists of vertices (9)--(16).}
\end{figure*}

The vertex model is also an important and intriguing topic in statistical mechanics. Studies of vertex models such as the six-vertex \cite{RN234, RN313, RN39, RN40, RN125, RN126, RN127, RN118, RN128, RN119, RN140, RN283}, eight-vertex \cite{RN65, RN59, RN287, RN60, RN54, RN193, RN196, RN197, RN198, RN61, RN194, RN129} and sixteen-vertex models \cite{RN56, RN63, RN130, RN131, RN132, RN124} are closely related to those of Ising models. The free-fermion models, which are solvable eight-vertex models that satisfy the free-fermion condition, can be exactly solved using various methods such as the \textit{S}-matrix \cite{RN65}, the Pfaffian \cite{RN59} and the transfer matrix methods \cite{RN314, RN315, RN316, RN317, RN318, RN319}. Typically, an eight-vertex model on the square lattice is defined as either the even or odd subcase of the general sixteen-vertex model \cite{RN124}, as shown in Fig.~\ref{fig1}. In this paper we adopt the notation of vertex configurations in Ref.~\cite{RN130} (see Fig.~1 there), and refer to the eight-vertex model of even (odd) subcase satisfying the free-fermion condition as even (odd) free-fermion model.

There have been lots of studies connecting the Ising model with the vertex model \cite{RN66, RN274, RN295, RN263, RN260, RN311}. Ref.~\cite{RN66} introduced a rather simple and direct correspondence between the square lattice Ising spin states and the arrow configurations of the sixteen-vertex model. The value $\pm1$ of an Ising spin is naturally mapped to the direction of the arrow settled on the corresponding site in the sixteen-vertex model. By further performing a transformation from the sixteen vertex weights to those in the even subcase, the author showed that the Ising model is equivalent with an even free-fermion model, both for the zero field case and the imginary field case. The present paper is inspired by the success of this work \cite{RN66}. The method of transforming the square lattice Ising model to the free-fermion model is extended and applied to the Ising models on the honeycomb, triangular and Kagomé lattices.

The structure of the paper is as follows. In Sec.~\ref{methods and techniques} we briefly introduce the methods and techniques used for transforming the Ising models into the free-fermion models, specifically the decorated lattice technique \cite{RN121, RN331, RN328}, the star-triangle transformation \cite{RN121, RN331, RN329, RN328, RN327} and the weak-graph expansion \cite{RN321, RN320, RN130, RN108}. In Sec.~\ref{results and discussions} we present the exact results of the Ising models on the honeycomb, triangular and Kagomé lattices. Each of these models is shown to be equivalent with a free-fermion model, both in cases under the zero field and the imaginary field. The exact solutions are then obtained in the free-fermion formulation. We summarize in Sec.~\ref{summary}.

\section{Methods}  \label{methods and techniques}
It is essential to introduce the methods and techniques for transformation of lattice systems and vertex weights, before we present the main results of the Ising models. The decorated lattice technique \cite{RN121, RN331, RN328} and the star-triangle transformation \cite{RN49, RN121, RN331, RN329, RN328, RN327} help us to obtain a lattice structure, which is convenient for translating the Boltzmann factors to the vertex weights of a sixteen-vertex model. The resulting sixteen-vertex model is then transformed into an even or odd eight-vertex model by the weak-graph expansion \cite{RN321, RN320, RN130, RN108}.

\subsection{Decorated lattice technique}
\begin{figure}
\includegraphics{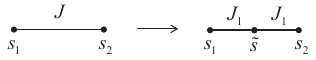}
\caption{\label{fig2}The diagrammatic representation of the decorated lattice technique. A spin $\tilde s$ is added and the interaction $J$ is replaced by $\tilde J$.}
\end{figure}

Decorated lattice technique is suitable for the case when we would like to add a spin site between a pair of nearest-neighbours, keeping the partition function invariant. As shown in Fig.~\ref{fig2}, the interaction $J$ is replaced by $\tilde J$. In our consideration, the additional spin $\tilde s$ is used to represent the direction of the arrow on the edge linking $s_1$ and $s_2$. This may help us to build up an effective vertex model formulation of the Ising problem. Consider the Boltzmann factors of this pair contributed to the partition function, which should be conserved after the decorated lattice transformation. That is 
\begin{equation}
e^{ - \beta J{s_1}{s_2}} = A\sum\limits_{\tilde s =  \pm 1} {{e^{ - \beta {\tilde J}\tilde s\left( {{s_1} + {s_2}} \right)}}}~.   \label{eq2}
\end{equation}
We can see clearly from Eq.~(\ref{eq2}) that
\begin{eqnarray*}
e^{ - \beta J} = A\left( {{e^{ - 2\beta {\tilde J}}} + {e^{2\beta {\tilde J}}}} \right), ~
e^{\beta J} = 2A. 
\end{eqnarray*}
Thus the interaction $\tilde J$ and the multiplier $A$ are determined by
\begin{equation}
\cosh \left( {2\beta {\tilde J}} \right) = e^{ - 2\beta J}~, \label{eq3}
\end{equation}
\begin{equation}
A = \frac{1}{2}{e^{\beta J}}~.    \label{eq4}
\end{equation}

\subsection{Star-triangle transformation}
\begin{figure}
\includegraphics{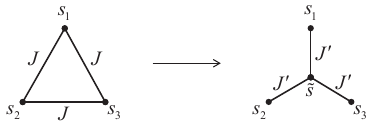}
\caption{\label{fig3}The diagrammatic representation of the star-triangle transformation. A spin $\tilde s$ is added and the interactions $J_1$, $J_2$ and $J_3$ are replaced by $J^{\prime}_1$, $J^{\prime}_2$ and $J^{\prime}_3$.}
\end{figure}

Fig.~\ref{fig3} shows the star-triangle transformation. The triangle with internal interactions $J_1$, $J_2$ and $J_3$ is replaced by a ``star'' consisting of an additional spin site coupled to the three vertices with interactions $J^{\prime}_1$, $J^{\prime}_2$ and $J^{\prime}_3$. The partition function should be invariant after the transformation, i.e., the Boltzmann factors of this unit should be conserved
\begin{equation}
e^{ - \beta \left( {{J_1}{s_1}{s_2} + {J_2}{s_2}{s_3} + {J_3}{s_3}{s_1}} \right)} = B\sum\limits_{\tilde s =  \pm 1} {e^{ - \beta \tilde s\left( {J^{\prime}_1{s_3} + J^{\prime}_2{s_1} + J^{\prime}_3{s_2}} \right)}}~. \label{eq5} 
\end{equation}
This leads to
\begin{align}
&{e^{ - \beta \left( {{J_1} + {J_2} + {J_3}} \right)}} = B\left[ {{e^{ - \beta \left( {J^{\prime}_1 + J^{\prime}_2 + J^{\prime}_3} \right)}} + {e^{ - \beta \left( {-J^{\prime}_1 - J^{\prime}_2 - J^{\prime}_3} \right)}}} \right],  \nonumber \\ 
&{e^{ - \beta \left( {{J_1} - {J_2} - {J_3}} \right)}} = B\left[ {{e^{ - \beta \left( { - J^{\prime}_1 + J^{\prime}_2 + J^{\prime}_3} \right)}} + {e^{ - \beta \left( {J^{\prime}_1 - J^{\prime}_2 - J^{\prime}_3} \right)}}} \right],  \nonumber \\ 
&{e^{ - \beta \left( { - {J_1} + {J_2} - {J_3}} \right)}} = B\left[ {{e^{ - \beta \left( {J^{\prime}_1 - J^{\prime}_2 + J^{\prime}_3} \right)}} + {e^{ - \beta \left( { - J^{\prime}_1 + J^{\prime}_2 - J^{\prime}_3} \right)}}} \right],  \nonumber \\ 
&{e^{ - \beta \left( { - {J_1} - {J_2} + {J_3}} \right)}} = B\left[ {{e^{ - \beta \left( {J^{\prime}_1 + J^{\prime}_2 - J^{\prime}_3} \right)}} + {e^{ - \beta \left( { - J^{\prime}_1 - J^{\prime}_2 + J^{\prime}_3} \right)}}} \right].  \nonumber 
\end{align}
The interactions $J^{\prime}_1$, $J^{\prime}_2$ and $J^{\prime}_3$ and the multiplier $B$ can be determined from the equations above \cite{RN49}
\begin{align}
&\cosh \left( 2\beta {J^{\prime}_1} \right) = \cosh \left( {2\beta {J_2}} \right)\cosh \left( {2\beta {J_3}} \right)   \nonumber \\
&~~~~~~~~~~~~~~~~~ - \sinh \left( {2\beta {J_2}} \right)\sinh \left( {2\beta {J_3}} \right)\coth \left( {2\beta {J_1}} \right),   \nonumber \\
&\cosh \left( 2\beta {J^{\prime}_2} \right) = \cosh \left( {2\beta {J_3}} \right)\cosh \left( {2\beta {J_1}} \right)   \nonumber \\
&~~~~~~~~~~~~~~~~~ - \sinh \left( {2\beta {J_3}} \right)\sinh \left( {2\beta {J_1}} \right)\coth \left( {2\beta {J_2}} \right),   \nonumber \\
&\cosh \left( 2\beta {J^{\prime}_3} \right) = \cosh \left( {2\beta {J_1}} \right)\cosh \left( {2\beta {J_2}} \right)   \nonumber \\
&~~~~~~~~~~~~~~~~~ - \sinh \left( {2\beta {J_1}} \right)\sinh \left( {2\beta {J_2}} \right)\coth \left( {2\beta {J_3}} \right),   \label{eq6}
\end{align}
\begin{align}
B =&\frac{{\left( {1 - t_1^2} \right)\left( {1 - t_2^2} \right)\left( {1 - t_3^2} \right)}}{4} \times  \nonumber \\
&\left[ {\frac{{ - \sinh \left( {2\beta {J_1}} \right)\sinh \left( {2\beta {J_2}} \right)\sinh \left( {2\beta {J_3}} \right)}}{{2\left( {1 - {t_1}{t_2}{t_3}} \right)\left( {{t_2}{t_3} - {t_1}} \right)\left( {{t_1}{t_3} - {t_2}} \right)\left( {{t_1}{t_2} - {t_3}} \right)}}} \right]^{{1 \mathord{\left/ {\vphantom {1 2}} \right. \kern-\nulldelimiterspace} 2}},   \label{eq7}
\end{align}
with ${t_i} = \tanh \left( {\beta {J_i}} \right) (i=1,2,3)$. One can easily verify that the star-triangle transformation directly connects the partition function of the triangular lattice model with that of the honeycomb lattice model. This enables us to simplify the derivation of the results for the triangular lattice model.  

\subsection{Weak-graph expansion}
Weak-graph expansion is a useful tool in lattice statistical problems \cite{RN321, RN108}. In this paper, we use the version of weak-graph expansion method proposed by Wu, which was originally designed for the even eight-vertex model \cite{RN320} and later extended to the sixteen-vertex model \cite{RN130}. Applying the method to the sixteen-vertex model enables us to obtain a new set of vertex weights, with the partition function invariant. The details of the method are shown in Ref.~\cite{RN320}, and we only list the result for all sixteen rearranged weights, expressed via a transformation matrix $\mathbf{M}$ as
$\left\{{{\tilde \omega }_1}, {{\tilde \omega }_2}, \ldots, {{\tilde \omega }_{16}}\right\}^T=\frac{1}{4}\mathbf{M}\left\{{{\omega }_1}, {{ \omega }_2}, \ldots, {{\omega }_{16}}\right\}^T$, with

\begin{widetext}
\begin{gather}
{\bf{M}} = \left( {\begin{array}{*{20}{c}}
1&1&1&1&1&1&1&1&1&1&1&1&1&1&1&1\\
1&1&1&1&1&1&1&1&{ - 1}&{ - 1}&{ - 1}&{ - 1}&{ - 1}&{ - 1}&{ - 1}&{ - 1}\\
1&1&1&1&{ - 1}&{ - 1}&{ - 1}&{ - 1}&{ - 1}&{ - 1}&{ - 1}&{ - 1}&1&1&1&1\\
1&1&1&1&{ - 1}&{ - 1}&{ - 1}&{ - 1}&1&1&1&1&{ - 1}&{ - 1}&{ - 1}&{ - 1}\\
1&1&{ - 1}&{ - 1}&1&1&{ - 1}&{ - 1}&{ - 1}&{ - 1}&1&1&1&1&{ - 1}&{ - 1}\\
1&1&{ - 1}&{ - 1}&1&1&{ - 1}&{ - 1}&1&1&{ - 1}&{ - 1}&{ - 1}&{ - 1}&1&1\\
1&1&{ - 1}&{ - 1}&{ - 1}&{ - 1}&1&1&{ - 1}&{ - 1}&1&1&{ - 1}&{ - 1}&1&1\\
1&1&{ - 1}&{ - 1}&{ - 1}&{ - 1}&1&1&1&1&{ - 1}&{ - 1}&1&1&{ - 1}&{ - 1}\\
{ - 1}&1&1&{ - 1}&1&{ - 1}&1&{ - 1}&1&{ - 1}&1&{ - 1}&{ - 1}&1&1&{ - 1}\\
{ - 1}&1&1&{ - 1}&1&{ - 1}&1&{ - 1}&{ - 1}&1&{ - 1}&1&1&{ - 1}&{ - 1}&1\\
{ - 1}&1&1&{ - 1}&{ - 1}&1&{ - 1}&1&1&{ - 1}&1&{ - 1}&1&{ - 1}&{ - 1}&1\\
{ - 1}&1&1&{ - 1}&{ - 1}&1&{ - 1}&1&{ - 1}&1&{ - 1}&1&{ - 1}&1&1&{ - 1}\\
{ - 1}&1&{ - 1}&1&{ - 1}&1&1&{ - 1}&{ - 1}&1&1&{ - 1}&1&{ - 1}&1&{ - 1}\\
{ - 1}&1&{ - 1}&1&{ - 1}&1&1&{ - 1}&1&{ - 1}&{ - 1}&1&{ - 1}&1&{ - 1}&1\\
{ - 1}&1&{ - 1}&1&1&{ - 1}&{ - 1}&1&1&{ - 1}&{ - 1}&1&1&{ - 1}&1&{ - 1}\\
{ - 1}&1&{ - 1}&1&1&{ - 1}&{ - 1}&1&{ - 1}&1&1&{ - 1}&{ - 1}&1&{ - 1}&1
\end{array}} \right).
 \label{eq8}
\end{gather}
\end{widetext}
The sixteen-vertex model is symmetric, when the weight of each vertex configuration is invariant under the reversal of arrows, i.e. 
\begin{equation}
{\omega _1} = {\omega _2},{\rm{ }} \cdots ,{\rm{ }}{\omega _{15}} = {\omega _{16}}.  \label{eq9}
\end{equation} 
Similarly, the antisymmetric sixteen-vertex model is one  satisfying
\begin{equation}
{\omega _1} = -{\omega _2},{\rm{ }} \cdots ,{\rm{ }}{\omega _{15}} = -{\omega _{16}}.  \label{eq10}
\end{equation} 
It is straightforward to see from Eq.~(\ref{eq8}) that, the symmetric sixteen-vertex model is transformed into an even eight-vertex model (Eq. (3) of Ref.~\cite{RN130} shows this case), while the antisymmetric case is transformed into an odd subcase.

We will show in the next section that each Ising model considered can be mapped to a symmetric/antisymmetric sixteen-vertex model, and further transformed into an even/odd eight-vertex model by Eq.~(\ref{eq8}). We can find that each case satisfies either the even free-fermion condition \cite{RN65, RN59}
\begin{equation}
{\tilde \omega _1}{\tilde \omega _2} + {\tilde \omega _3}{\tilde \omega _4} = {\tilde \omega _5}{\tilde \omega _6} + {\tilde \omega _7}{\tilde \omega _8}  \label{eq11}
\end{equation}
or the odd free-fermion condition \cite{RN129}
\begin{equation}
{\tilde \omega _9}{\tilde \omega _{10}} + {\tilde \omega _{11}}{\tilde \omega _{12}} = {\tilde \omega _{13}}{\tilde \omega _{14}} + {\tilde \omega _{15}}{\tilde \omega _{16}}~.   \label{eq12}
\end{equation}

\section{Results}   \label{results and discussions}
\subsection{The honeycomb lattice}
\begin{figure} 
 \includegraphics{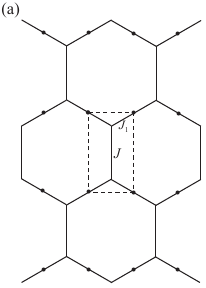}
 \quad \quad
 \includegraphics{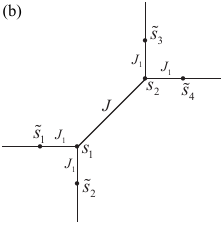}
 \caption{The diagrammatic representation of applying the decorated lattice technique to the honeycomb lattice model. The region bounded by dashed lines in (a) forms a vertex site of the sixteen-vertex model. The details of the vertex unit are shown in (b).} \label{fig4}
\end{figure}

We apply the decorated lattice technique to the honeycomb lattice, as shown in Fig.~\ref{fig4}(a). It is easy to verify that, the region bounded by dash lines can be regarded as a vertex site on the square lattice. Each vertex site consists of two honeycomb lattice Ising spins. Fig.~\ref{fig4}(b) shows the details with $\tilde J_2$ and $\tilde J_3$ determined from $J_2$ and $J_3$ by Eq.~(\ref{eq3}), respectively. It is natural to introduce a one-to-one mapping from the spin states of $\left\{ {\tilde s} \right\}$ in Fig.~\ref{fig4}(b) to the arrow configurations
\begin{eqnarray}
&&\text{for horizontal edge} ~~ \mathop { \frac{~~}{~~} \!\!\bullet\!\! \frac{~~}{~~}} ~+/- ~\Rightarrow~ \rightarrow/ \leftarrow, \nonumber \\  
&&\text{for vertical edge} ~~ \mathop {\bullet} \limits_{|}^{|} ~+/- ~\Rightarrow~ \uparrow/ \downarrow.  \label{eq13}
\end{eqnarray}
Therefore, a sixteen-vertex model can be built up with an appropriate correspondence between the vertex weights and the Boltzmann factors of the Ising model. 

\subsubsection{Zero-field case}
In the zero-field case, the partition function can be expressed by
\begin{equation}
Z = A_2^{N/2}A_3^{N/2} \sum\limits_{\left\{ {\tilde s} \right\} =  \pm 1}~{\prod\limits_{{\rm{all~vertices}}} {\omega \left( {{{\tilde s}_1},{{\tilde s}_2},{{\tilde s}_3},{{\tilde s}_4}} \right)} }   \label{eq14}
\end{equation}
with
\begin{align}
&\omega \left( {{{\tilde s}_1},{{\tilde s}_2},{{\tilde s}_3},{{\tilde s}_4}} \right) = \sum\limits_{{s_1} = \pm 1,{s_2} =  \pm 1} e^{ - \beta J_1{s_1}{s_2}}\times  \nonumber  \\
&~~~~~~~~~~~~~~~~~~~~~~~~~~e^{ - \beta \left[ {{{\tilde J}_2}\left( {{s_1}{{\tilde s}_2} + {s_2}{{\tilde s}_3}} \right) + {{\tilde J}_3}\left( {{s_1}{{\tilde s}_1} + {s_2}{{\tilde s}_4}} \right)} \right]}   \label{eq15}
\end{align}
and $A_i$ determined from $J_i$ by Eq.~(\ref{eq4}). $\omega \left( {{{\tilde s}_1},{{\tilde s}_2},{{\tilde s}_3},{{\tilde s}_4}} \right)$ is the Boltzmann factor of the vertex unit shown in Fig.~\ref{fig4}(b). Then the vertex weights can be determined directly from $\omega \left( {{{\tilde s}_1},{{\tilde s}_2},{{\tilde s}_3},{{\tilde s}_4}} \right)$, according to the vertex configurations in Fig.~\ref{fig1} and the mapping in Eq.~(\ref{eq13}). That is 
\begin{align}
&\omega _1 = \omega \left( { + + + + } \right) = 2{e^{\beta {J_1}}} + {e^{ - \beta {J_1}}}\left[ {{e^{2\beta \left( {{{\tilde J}_2} + {{\tilde J}_3}} \right)}} + {e^{ - 2\beta \left( {{{\tilde J}_2} + {{\tilde J}_3}} \right)}}} \right],  \nonumber  \\
& \omega _2 = \omega _1,  \nonumber \\
& {\omega _3} = \omega \left( { + - - + } \right) = 2{e^{\beta {J_1}}} + {e^{ - \beta {J_1}}}\left[ {{e^{2\beta \left( {{{\tilde J}_2} - {{\tilde J}_3}} \right)}} + {e^{ - 2\beta \left( {{{\tilde J}_2} - {{\tilde J}_3}} \right)}}} \right],  \nonumber \\
& \omega _4 = \omega _3,  \nonumber \\
& {\omega _5} = \omega \left( { + - + - } \right) = 2{e^{-\beta {J_1}}} + {e^{ \beta {J_1}}}\left[ {{e^{2\beta \left( {{{\tilde J}_2} - {{\tilde J}_3}} \right)}} + {e^{ - 2\beta \left( {{{\tilde J}_2} - {{\tilde J}_3}} \right)}}} \right],  \nonumber \\
& \omega _6 = \omega _5,  \nonumber \\
& \omega _7 = \omega \left( { - - + + } \right) = 2{e^{-\beta {J_1}}} + {e^{ \beta {J_1}}}\left[ {{e^{2\beta \left( {{{\tilde J}_2} + {{\tilde J}_3}} \right)}} + {e^{ - 2\beta \left( {{{\tilde J}_2} + {{\tilde J}_3}} \right)}}} \right],  \nonumber \\
& \omega _8 = \omega _7,  \nonumber \\
& {\omega _9} = \omega \left( { + - + + } \right)  \nonumber  \\
&~~~~ ={e^{\beta {J_1}}}\left( {{e^{2\beta {{\tilde J}_2}}} + {e^{ - 2\beta {{\tilde J}_2}}}} \right) + {e^{ - \beta {J_1}}}\left( {{e^{2\beta {{\tilde J}_3}}} + {e^{ - 2\beta {{\tilde J}_3}}}} \right),  \nonumber \\
& \omega _{10} = \omega _{11}  = \omega _{12} = \omega _9,  \nonumber \\
& \omega _{13} = \omega \left( { - + + + } \right)  \nonumber  \\
&~~~~~ ={e^{-\beta {J_1}}}\left( {{e^{2\beta {{\tilde J}_2}}} + {e^{ - 2\beta {{\tilde J}_2}}}} \right) + {e^{ \beta {J_1}}}\left( {{e^{2\beta {{\tilde J}_3}}} + {e^{ - 2\beta {{\tilde J}_3}}}} \right),  \nonumber \\
&\omega _{14} = \omega _{15}  = \omega _{16} = \omega _{13}. 
\label{eq16}
\end{align}
We see that the resulting sixteen-vertex model is symmetric. As stated before, the vertex weights can be rearranged by performing the weak-graph expansion such that an even eight-vertex model is obtained. The new weights are straightforward to show from Eq.~(\ref{eq8}):
\begin{align}
{\tilde \omega }_1  &  = \frac{1}{2}{\omega _1} + \frac{1}{2}{\omega _3} + \frac{1}{2}{\omega _5} + \frac{1}{2}{\omega _7} + {\omega _9} + {\omega _{13}}   \nonumber  \\
& = 2\left( {{e^{\beta {J_1}}} + {e^{ - \beta {J_1}}}} \right)\left( {{e^{ - 2\beta {{J}_2}}} + 1} \right)\left( {{e^{ - 2\beta {{J}_3}}} + 1} \right),  \nonumber  \\
 {\tilde \omega }_2 &  = \frac{1}{2}{\omega _1} + \frac{1}{2}{\omega _3} + \frac{1}{2}{\omega _5} + \frac{1}{2}{\omega _7} - {\omega _9} - {\omega _{13}}   \nonumber  \\
& = 2\left( {{e^{\beta {J_1}}} + {e^{ - \beta {J_1}}}} \right)\left( {{e^{ - 2\beta {{J}_2}}} - 1} \right)\left( {{e^{ - 2\beta {{J}_3}}} - 1} \right),  \nonumber  \\
 {\tilde \omega }_3  & = \frac{1}{2}{\omega _1} + \frac{1}{2}{\omega _3} - \frac{1}{2}{\omega _5} - \frac{1}{2}{\omega _7} - {\omega _9} + {\omega _{13}}   \nonumber  \\
& = 2\left( {{e^{\beta {J_1}}} - {e^{ - \beta {J_1}}}} \right)\left( {-{e^{ - 2\beta {{J}_2}}} + 1} \right)\left( {{e^{ - 2\beta {{J}_3}}} + 1} \right),  \nonumber  \\
 {\tilde \omega }_4  & = \frac{1}{2}{\omega _1} + \frac{1}{2}{\omega _3} - \frac{1}{2}{\omega _5} - \frac{1}{2}{\omega _7} + {\omega _9} - {\omega _{13}}   \nonumber  \\
& = 2\left( {{e^{\beta {J_1}}} - {e^{ - \beta {J_1}}}} \right)\left( {{e^{ - 2\beta {{J}_2}}} + 1} \right)\left( {-{e^{ - 2\beta {{J}_3}}} + 1} \right),  \nonumber  \\
 {\tilde \omega }_5  & = \frac{1}{2}{\omega _1} - \frac{1}{2}{\omega _3} + \frac{1}{2}{\omega _5} - \frac{1}{2}{\omega _7}  \nonumber  \\
& = -2\left( {{e^{\beta {J_1}}} - {e^{ - \beta {J_1}}}} \right) \left( {{e^{ - 4\beta {{J}_2}}} - 1} \right)^{1/2} \left( {{e^{ - 4\beta {{J}_3}}} - 1} \right)^{1/2},  \nonumber  \\
 {\tilde \omega }_6 & = {\tilde \omega }_5,  \nonumber  \\
 {\tilde \omega }_7 & = \frac{1}{2}{\omega _1} - \frac{1}{2}{\omega _3} - \frac{1}{2}{\omega _5} + \frac{1}{2}{\omega _7}  \nonumber  \\
& = 2\left( {{e^{\beta {J_1}}} + {e^{ - \beta {J_1}}}} \right) \left( {{e^{ - 4\beta {{J}_2}}} - 1} \right)^{1/2} \left( {{e^{ - 4\beta {{J}_3}}} - 1} \right)^{1/2},  \nonumber  \\
{\tilde \omega }_8  & = {\tilde \omega }_7. 
\label{eq17}
\end{align}

One can easily examine that the even free-fermion condition Eq.~(\ref{eq11}) is satisfied. Thus, the honeycomb lattice Ising model in the zero field is equivalent with an even free-fermion model. Denote the number of vertex sites by $N_{\rm{v}}$. The exact solution of the even free-fermion model $\mathop {\lim }\limits_{{N_{{\rm{v}}}} \to \infty } \frac{1}{{{N_{{\rm{v}}}}}}\ln {Z_{{\rm{even}}}}$ had been given in Refs.~\cite{RN65, RN59} (note that the ordering of vertices (5)--(8) in Fig.~\ref{fig1} differs slightly from that of Refs.~\cite{RN65, RN59}, but the expression for the solution is the same). We show the result quoting this well-known solution and noticing Eq.~(\ref{eq14})
\begin{align}
\mathop {\lim }\limits_{N \to \infty } \frac{1}{N}\ln Z  = \frac{1}{2} ( \ln A_2 + \ln A_3 ) + \frac{1}{2}\mathop {\lim }\limits_{{N_{{\rm{v}}}} \to \infty } \frac{1}{{{N_{{\rm{v}}}}}}\ln {Z_{{\rm{even}}}}  \nonumber \\
= \frac{1}{2} \ln \left( {\frac{1}{2}{e^{\beta J_2}}} \right) + \frac{1}{2} \ln \left( {\frac{1}{2}{e^{\beta J_3}}} \right) + \frac{1}{2} \frac{1}{{8{\pi ^2}}}\int_0^{2\pi } d\theta \int_0^{2\pi } d\phi   \nonumber \\
~~~~\ln \left[ a + b\cos \theta + c\cos \phi + d\cos \left( {\theta  - \phi } \right) + e\cos \left( {\theta  + \phi }\right) \right]   \label{eq18}
\end{align}
with
\begin{eqnarray}
a &=& \tilde \omega _1^2 + \tilde \omega _2^2 + \tilde \omega _3^2 + \tilde \omega _4^2,  \nonumber \\
b &=& 2({{\tilde \omega }_1}{{\tilde \omega }_3} - {{\tilde \omega }_2}{{\tilde \omega }_4}),  \nonumber \\
c &=& 2({{\tilde \omega }_1}{{\tilde \omega }_4} - {{\tilde \omega }_2}{{\tilde \omega }_3}), \nonumber \\
d &=& 2({{\tilde \omega }_3}{{\tilde \omega }_4} - {{\tilde \omega }_7}{{\tilde \omega }_8}), \nonumber \\
e &=& 2({{\tilde \omega }_3}{{\tilde \omega }_4} - {{\tilde \omega }_5}{{\tilde \omega }_6}). \label{eq19}
\end{eqnarray}
A straightforward calculation of the coefficients in Eq.~(\ref{eq19}) from the vertex weights in Eq.~(\ref{eq17}) gives
\begin{align}
&a = {2^7}e^{ - 2\beta (J_2 + J_3)}\left[ \cosh\left( 2\beta J_1 \right) \cosh\left( 2\beta J_2 \right) \cosh\left( 2\beta J_3 \right) + 1 \right],  \nonumber \\
&b = {2^7}e^{ - 2\beta (J_2 + J_3)} \sinh\left( 2\beta J_1 \right) \sinh\left( 2\beta J_2 \right),  \nonumber \\
&c = {2^7}e^{ - 2\beta (J_2 + J_3)} \sinh\left( 2\beta J_1 \right) \sinh\left( 2\beta J_3 \right),  \nonumber \\
&d = -{2^7}e^{ - 2\beta (J_2 + J_3)} \sinh\left( 2\beta J_2 \right) \sinh\left( 2\beta J_3 \right),  \nonumber \\
&e = 0.  \label{eq20}
\end{align}
Inserting Eq.~(\ref{eq20}) into Eq.~(\ref{eq18}) and making some arrangements, the exact solution turns out to be
\begin{align}
&\mathop {\lim }\limits_{N \to \infty } \frac{1}{N}\ln Z = \frac{3}{4}\ln 2 + \frac{1}{{16{\pi ^2}}}\int_0^{2\pi } d\theta \int_0^{2\pi } d\phi   \nonumber \\
&~~~~~~~~~~~\ln \left[ \cosh\left( 2\beta J_1 \right) \cosh\left( 2\beta J_2 \right) \cosh\left( 2\beta J_3 \right) + 1 \right.  \nonumber \\
&~~~~~~~~~~~~~~~+ \sinh\left( 2\beta J_1 \right) \sinh\left( 2\beta J_2 \right) \cos \theta  \nonumber \\
&~~~~~~~~~~~~~~~+ \sinh\left( 2\beta J_1 \right) \sinh\left( 2\beta J_3 \right) \cos \phi   \nonumber \\
&~~~~~~~~~~~~~~~\left. - \sinh\left( 2\beta J_2 \right) \sinh\left( 2\beta J_3 \right) \cos \left( \theta  - \phi  \right) \right].  \label{eq21}
\end{align}
One can verify that, this expression is consistent with the results published in the previous literatures \cite{RN122, RN123, RN221, RN224}. 

\subsubsection{Imaginary-field case}
In the case of $H_{\rm{ex}}=i( {{\pi  \mathord{\left/{\vphantom {\pi  2}} \right. \kern-\nulldelimiterspace} 2}} ){k_B}T$, we notice that
\begin{equation}
{e^{\beta {H_{{\rm{ex}}}}\sum\limits_{i = 1}^N {{s_i}} }} = {i^{\sum\limits_{i = 1}^N {{s_i}} }} = {i^N}\prod\limits_{i = 1}^N {{s_i}}  \label{eq22}
\end{equation}
by using the identity ${i^{{s_i}}} = i \times {s_i}$ \cite{RN274}. We can see that, $i^N$ is a constant and the product $\prod\limits_{i = 1}^N {{s_i}}$ can be conveniently included in the factors $\omega \left( {{{\tilde s}_1},{{\tilde s}_2},{{\tilde s}_3},{{\tilde s}_4}} \right)$. We just need to rewrite the partition function as
\begin{equation}
Z = i^N A_2^{N/2}A_3^{N/2} \sum\limits_{\left\{ {\tilde s} \right\} =  \pm 1}~{\prod\limits_{{\rm{all~vertices}}} {\omega \left( {{{\tilde s}_1},{{\tilde s}_2},{{\tilde s}_3},{{\tilde s}_4}} \right)} }   \label{eq23}
\end{equation}
with 
\begin{align}
&\omega \left( {{{\tilde s}_1},{{\tilde s}_2},{{\tilde s}_3},{{\tilde s}_4}} \right) = \sum\limits_{{s_1} = \pm 1,{s_2} =  \pm 1} (s_1 s_2) e^{ - \beta J_1{s_1}{s_2}}\times  \nonumber  \\
&~~~~~~~~~~~~~~~~~~~~~~~~~{{e^{ - \beta \left[ {{{\tilde J}_2}\left( {{s_1}{{\tilde s}_2} + {s_2}{{\tilde s}_3}} \right) + {{\tilde J}_3}\left( {{s_1}{{\tilde s}_1} + {s_2}{{\tilde s}_4}} \right)} \right]}}}.  \label{eq24}
\end{align}
The vertex weights are then determined in the similar way to those in Eq.~(\ref{eq16})
\begin{align}
&\omega _1 = \omega _2 = -2{e^{\beta {J_1}}} + {e^{ - \beta {J_1}}}\left[ {{e^{2\beta \left( {{{\tilde J}_2} + {{\tilde J}_3}} \right)}} + {e^{ - 2\beta \left( {{{\tilde J}_2} + {{\tilde J}_3}} \right)}}} \right],  \nonumber  \\
&\omega _3 = \omega _4 = -2{e^{\beta {J_1}}} + {e^{ - \beta {J_1}}}\left[ {{e^{2\beta \left( {{{\tilde J}_2} - {{\tilde J}_3}} \right)}} + {e^{ - 2\beta \left( {{{\tilde J}_2} - {{\tilde J}_3}} \right)}}} \right],  \nonumber \\
&\omega _5 = \omega _6 = 2{e^{-\beta {J_1}}} - {e^{ \beta {J_1}}}\left[ {{e^{2\beta \left( {{{\tilde J}_2} - {{\tilde J}_3}} \right)}} + {e^{ - 2\beta \left( {{{\tilde J}_2} - {{\tilde J}_3}} \right)}}} \right],  \nonumber \\
&\omega _7 = \omega _8 = 2{e^{-\beta {J_1}}} - {e^{ \beta {J_1}}}\left[ {{e^{2\beta \left( {{{\tilde J}_2} + {{\tilde J}_3}} \right)}} + {e^{ - 2\beta \left( {{{\tilde J}_2} + {{\tilde J}_3}} \right)}}} \right],  \nonumber \\
&\omega _9 = \omega _{10} = \omega _{11}  = \omega _{12}   \nonumber  \\
&~~~={-e^{\beta {J_1}}}\left( {{e^{2\beta {{\tilde J}_2}}} + {e^{ - 2\beta {{\tilde J}_2}}}} \right) + {e^{ - \beta {J_1}}}\left( {{e^{2\beta {{\tilde J}_3}}} + {e^{ - 2\beta {{\tilde J}_3}}}} \right),  \nonumber \\
&\omega _{13} = \omega _{14} = \omega _{15}  = \omega _{16}  \nonumber  \\
&~~~~~={e^{-\beta {J_1}}}\left( {{e^{2\beta {{\tilde J}_2}}} + {e^{ - 2\beta {{\tilde J}_2}}}} \right) - {e^{ \beta {J_1}}}\left( {{e^{2\beta {{\tilde J}_3}}} + {e^{ - 2\beta {{\tilde J}_3}}}} \right). 
\label{eq25}
\end{align}
Again, the Boltzmann factors are translated into the weights of a symmetric sixteen-vertex model. We still perform the weak-graph expansion via Eq.~(\ref{eq8})
\begin{align}
{\tilde \omega }_1 &= 2\left( {-{e^{\beta {J_1}}} + {e^{ - \beta {J_1}}}} \right)\left( {{e^{ - 2\beta {{J}_2}}} + 1} \right)\left( {{e^{ - 2\beta {{J}_3}}} + 1} \right),  \nonumber  \\
{\tilde \omega }_2 &= 2\left( {-{e^{\beta {J_1}}} + {e^{ - \beta {J_1}}}} \right)\left( {{e^{ - 2\beta {{J}_2}}} - 1} \right)\left( {{e^{ - 2\beta {{J}_3}}} - 1} \right),  \nonumber  \\
{\tilde \omega }_3 &= -2\left( {{e^{\beta {J_1}}} + {e^{ - \beta {J_1}}}} \right)\left( {-{e^{ - 2\beta {{J}_2}}} + 1} \right)\left( {{e^{ - 2\beta {{J}_3}}} + 1} \right),  \nonumber  \\
{\tilde \omega }_4 &= -2\left( {{e^{\beta {J_1}}} + {e^{ - \beta {J_1}}}} \right)\left( {{e^{ - 2\beta {{J}_2}}} + 1} \right)\left( {-{e^{ - 2\beta {{J}_3}}} + 1} \right),  \nonumber  \\
{\tilde \omega }_5 &= {\tilde \omega }_6    \nonumber  \\
&= 2\left( {{e^{\beta {J_1}}} + {e^{ - \beta {J_1}}}} \right) \left( {{e^{ - 4\beta {{J}_2}}} - 1} \right)^{1/2} \left( {{e^{ - 4\beta {{J}_3}}} - 1} \right)^{1/2},  \nonumber  \\
{\tilde \omega }_7 &= {\tilde \omega }_8   \nonumber  \\
&=2\left( {-{e^{\beta {J_1}}} + {e^{ - \beta {J_1}}}} \right) \left( {{e^{ - 4\beta {{J}_2}}} - 1} \right)^{1/2} \left( {{e^{ - 4\beta {{J}_3}}} - 1} \right)^{1/2}. 
\label{eq26}
\end{align}
Clearly, we have obtained an even free-fermion model again. Notice Eq.~(\ref{eq23}) and the partition function is
\begin{equation}
\mathop {\lim }\limits_{N \to \infty } \frac{1}{N}\ln Z = i\frac{\pi }{2} +  \frac{1}{2}( \ln A_2 + \ln A_3 ) + \frac{1}{2}\mathop {\lim }\limits_{N_{\rm{v}} \to \infty } \frac{1}{N_{\rm{v}}}\ln Z_{\rm{even}}   \label{eq27}
\end{equation}
with the even free-fermion solution expressed in Eqs.~(\ref{eq18}) and (\ref{eq19}). Omitting the detailed calculations, we show the final exact result 
\begin{align}
& \mathop {\lim }\limits_{N \to \infty } \frac{1}{N}\ln Z = i\frac{\pi }{2} + \frac{3}{4}\ln 2 + \frac{1}{{16{\pi ^2}}}\int_0^{2\pi } d\theta \int_0^{2\pi } d\phi   \nonumber \\
&~~~~~~~~~~~ \ln \left[ \cosh\left( 2\beta J_1 \right) \cosh\left( 2\beta J_2 \right) \cosh\left( 2\beta J_3 \right) - 1 \right.  \nonumber \\
&~~~~~~~~~~~~~~~ + \sinh\left( 2\beta J_1 \right) \sinh\left( 2\beta J_2 \right) \cos \theta   \nonumber \\
&~~~~~~~~~~~~~~~ + \sinh\left( 2\beta J_1 \right) \sinh\left( 2\beta J_3 \right) \cos \phi   \nonumber \\
&~~~~~~~~~~~~~~~\left. + \sinh\left( 2\beta J_2 \right) \sinh\left( 2\beta J_3 \right) \cos \left( \theta  - \phi  \right) \right].  \label{eq28}
\end{align}
One can examine that our result recovers the proposed solutions of this case \cite{RN67, RN274}.

\subsection{The triangular lattice}
\begin{figure} 
 \includegraphics{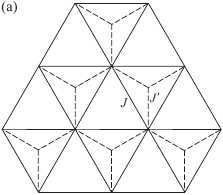}
 \quad
 \includegraphics{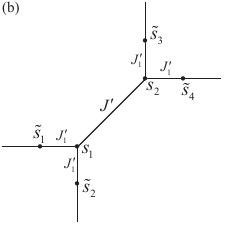}
 \caption{The diagrammatic representation of applying the star-triangle transformation to the triangular lattice model. The transformed system, which is on the honeycomb lattice as shown by the dash lines in (a), can be mapped into the sixteen-vertex model in the same way as that in Fig.~4(a). The details of the vertex unit are shown in (b).} \label{fig5}
\end{figure}

To deal with the triangular lattice Ising model, we employ the star-triangle transformation as shown in Fig.~\ref{fig5}(a). The transformed spin system is on the honeycomb lattice, with the replacement of the nearest-neighbour interactions $J_i \to J^{\prime}_i$. The new anisotropic interactions $J^{\prime}_i$ are determined by Eq.~(\ref{eq6}). Applying the decorated lattice technique in the same way as that in Fig.~\ref{fig4}(a), we construct a sixteen-vertex model of which the vertex site is shown Fig.~\ref{fig5}(b). The mapping from the spin states to the arrow configurations is still that in Eq.~(\ref{eq13}). The relation between $J^{\prime}_i$ and ${\tilde J'}_i$ is the same with that in Eq.~(\ref{eq3}). 

\subsubsection{Zero-field case}
In the zero field, the partition function can be obtained using the result for the honeycomb lattice as: 
\begin{eqnarray*}
Z = {B^N}Z\left( {\rm{honeycomb}},J_i \to J^{\prime}_i,H_{\rm{ex}}=0 \right) 
\end{eqnarray*}
with $B$ defined in Eq.~\eqref{eq7}. Obviously this case is also equivalent with an even free-fermion model. Since the number of spins on the transformed honeycomb lattice is double that on the triangular lattice, the partition function should be
\begin{equation}
\mathop {\lim }\limits_{N \to \infty } \frac{1}{N}\ln Z = \ln B + 2 \times {\rm{Eq}}.~\eqref{eq21}\left( J_i \to J^{\prime}_i \right).  \label{eq29}
\end{equation}
Taking into account Eqs.~(\ref{eq6}), (\ref{eq7}) and (\ref{eq21}), the exact solution can be obtained 
\begin{align}
& \mathop {\lim }\limits_{N \to \infty } \frac{1}{N}\ln Z = \ln2 + \frac{1}{{8{\pi ^2}}}\int_0^{2\pi } d\theta \int_0^{2\pi } d\phi  \nonumber \\
& \ln \left[ \cosh\left( 2\beta J_1 \right) \cosh\left( 2\beta J_2 \right) \cosh\left( 2\beta J_3 \right) \right.  \nonumber \\
&~ - \sinh\left( 2\beta J_1 \right) \sinh\left( 2\beta J_2 \right) \sinh\left( 2\beta J_3 \right)  - \sinh\left( 2\beta J_3 \right) \cos \theta  \nonumber \\
&~ \left. - \sinh\left( 2\beta J_2 \right) \cos \phi + \sinh\left( 2\beta J_1 \right) \cos \left( \theta  - \phi  \right) \right].  \label{eq30}
\end{align}
This expression agrees with the known results of this case \cite{RN81, RN220, RN299, RN223}.

\subsubsection{Imaginary-field case}
We still make use of Eq.~(\ref{eq22}) to deal with the effect of the imaginary field. The partition function of this case should be 
\begin{equation}
Z = i^N B^N {A'_2}^N {A'_3}^N \sum\limits_{\left\{ {\tilde s} \right\} =  \pm 1}~{\prod\limits_{{\rm{all~vertices}}} {\omega \left( {{{\tilde s}_1},{{\tilde s}_2},{{\tilde s}_3},{{\tilde s}_4}} \right)} }  \label{eq31}
\end{equation}
with $A'_i = \frac{1}{2}{e^{\beta J'_i}}$ and
\begin{align}
&\omega \left( {{{\tilde s}_1},{{\tilde s}_2},{{\tilde s}_3},{{\tilde s}_4}} \right) = \sum\limits_{{s_1} = \pm 1,{s_2} =  \pm 1} (s_1)e^{ - \beta J'_1{s_1}{s_2}}\times  \nonumber  \\
&~~~~~~~~~~~~~~~~~~~~~~~~~e^{ - \beta \left[ {{{\tilde J'}_2}\left( {{s_1}{{\tilde s}_2} + {s_2}{{\tilde s}_3}} \right) + {{\tilde J'}_3}\left( {{s_1}{{\tilde s}_1} + {s_2}{{\tilde s}_4}} \right)} \right]}.  \label{eq32}
\end{align}
Note that the factors $\omega \left( {{{\tilde s}_1},{{\tilde s}_2},{{\tilde s}_3},{{\tilde s}_4}} \right)$ differ from that in Eq.~(\ref{eq24}) as the spin $s_2$ is not on the original triangular lattice. Similarly, the factors $\omega \left( {{{\tilde s}_1},{{\tilde s}_2},{{\tilde s}_3},{{\tilde s}_4}} \right)$ are translated into the vertex weights of a sixteen-vertex model. We find that the vertex weights in this case are antisymmetric 
\begin{align}
&\omega _1 = {e^{ - \beta {J^{\prime}_1}}}\left[ {{e^{ - 2\beta \left( {{{\tilde J'}_2} + {{\tilde J'}_3}} \right)}} - {e^{2\beta \left( {{{\tilde J'}_2} + {{\tilde J'}_3}} \right)}}} \right],  \nonumber \\
&\omega _2 = -\omega _1,  \nonumber \\
&\omega _3 = {e^{ - \beta {J^{\prime}_1}}}\left[ {{e^{ 2\beta \left( {{{\tilde J'}_2} - {{\tilde J'}_3}} \right)}} - {e^{-2\beta \left( {{{\tilde J'}_2} - {{\tilde J'}_3}} \right)}}} \right],  \nonumber \\
&\omega _4 = -\omega _3,  \nonumber \\
&\omega _5 = {e^{\beta {J^{\prime}_1}}}\left[ {{e^{ 2\beta \left( {{{\tilde J'}_2} - {{\tilde J'}_3}} \right)}} - {e^{-2\beta \left( {{{\tilde J'}_2} - {{\tilde J'}_3}} \right)}}} \right],  \nonumber \\
&\omega _6 = -\omega _5,  \nonumber \\
&\omega _7 = {e^{\beta {J^{\prime}_1}}}\left[ {{e^{ 2\beta \left( {{{\tilde J'}_2} + {{\tilde J'}_3}} \right)}} - {e^{-2\beta \left( {{{\tilde J'}_2} + {{\tilde J'}_3}} \right)}}} \right],  \nonumber \\
&\omega _8 = -\omega _7,  \nonumber \\
&\omega _9 = e^{\beta J^{\prime}_1}\left( {{e^{ 2\beta {\tilde J'}_2}} - {e^{-2\beta {\tilde J'}_2 }}} \right) + e^{-\beta J^{\prime}_1}\left( e^{ -2\beta {\tilde J'}_3} - e^{ 2\beta {\tilde J'}_3} \right),  \nonumber \\
&\omega _{10} = -\omega _9,  \nonumber \\
&\omega _{11} = e^{\beta J^{\prime}_1}\left( {{e^{ 2\beta {\tilde J'}_2}} - {e^{-2\beta {\tilde J'}_2 }}} \right) + e^{-\beta J^{\prime}_1}\left( e^{ 2\beta {\tilde J'}_3} - e^{ -2\beta {\tilde J'}_3} \right),  \nonumber \\
&\omega _{12} = -\omega _{11},  \nonumber \\
&\omega _{13} = e^{-\beta J^{\prime}_1}\left( {{e^{ -2\beta {\tilde J'}_2}} - {e^{2\beta {\tilde J'}_2 }}} \right) + e^{\beta J^{\prime}_1}\left( e^{ 2\beta {\tilde J'}_3} - e^{ -2\beta {\tilde J'}_3} \right),  \nonumber \\
&\omega _{14} = -\omega _{13},  \nonumber \\
&\omega _{15} = e^{-\beta J^{\prime}_1}\left( e^{ 2\beta {\tilde J'}_2} - e^{ -2\beta {\tilde J'}_2} \right) + e^{\beta J^{\prime}_1}\left( {{e^{ 2\beta {\tilde J'}_3}} - {e^{-2\beta {\tilde J'}_3}}} \right),  \nonumber \\
&\omega _{16} = -\omega _{15}.  \label{eq33}
\end{align}
From these weights, the weak-graph expansion produces an odd eight-vertex model by Eq.~(\ref{eq8})
\begin{align}
&\tilde \omega_9 = 2\cosh \left( \beta J^{\prime}_1 \right)\left\{ \sinh \left[ {2\beta \left( {\tilde J'}_2 + {\tilde J'}_3 \right)} \right]   \right.  \nonumber  \\
&~~~~~~~~~~~~~~~~~~\left. + \sinh \left[ {2\beta \left( {\tilde J'}_2 - {\tilde J'}_3 \right)} \right] + 2\sinh \left( {2\beta {{\tilde J'}_2}} \right) \right\},  \nonumber  \\
&\tilde \omega_{10} = 2\cosh \left( \beta J^{\prime}_1 \right)\left\{ \sinh \left[ {2\beta \left( {\tilde J'}_2 + {\tilde J'}_3 \right)} \right]   \right.  \nonumber  \\
&~~~~~~~~~~~~~~~~~~\left. + \sinh \left[ {2\beta \left( {\tilde J'}_2 - {\tilde J'}_3 \right)} \right] - 2\sinh \left( {2\beta {{\tilde J'}_2}} \right) \right\},  \nonumber  \\
&\tilde \omega_{11} = -2\sinh \left( \beta J^{\prime}_1 \right)\left\{ \sinh \left[ {2\beta \left( {\tilde J'}_2 + {\tilde J'}_3 \right)} \right]   \right.  \nonumber  \\
&~~~~~~~~~~~~~~~~~~\left. + \sinh \left[ {2\beta \left( {\tilde J'}_2 - {\tilde J'}_3 \right)} \right] - 2\sinh \left( {2\beta {{\tilde J'}_2}} \right) \right\},  \nonumber  \\
&\tilde \omega_{12} = -2\sinh \left( \beta J^{\prime}_1 \right)\left\{ \sinh \left[ {2\beta \left( {\tilde J'}_2 + {\tilde J'}_3 \right)} \right]   \right.  \nonumber  \\
&~~~~~~~~~~~~~~~~~~\left. + \sinh \left[ {2\beta \left( {\tilde J'}_2 - {\tilde J'}_3 \right)} \right] + 2\sinh \left( {2\beta {{\tilde J'}_2}} \right) \right\},  \nonumber  \\
&\tilde \omega_{13} = 2\cosh \left( \beta J^{\prime}_1 \right)\left\{ \sinh \left[ {2\beta \left( {\tilde J'}_2 + {\tilde J'}_3 \right)} \right]   \right.  \nonumber  \\
&~~~~~~~~~~~~~~~~~~\left. - \sinh \left[ {2\beta \left( {\tilde J'}_2 - {\tilde J'}_3 \right)} \right] + 2\sinh \left( {2\beta {{\tilde J'}_3}} \right) \right\},  \nonumber  \\
&\tilde \omega_{14} = 2\cosh \left( \beta J^{\prime}_1 \right)\left\{ \sinh \left[ {2\beta \left( {\tilde J'}_2 + {\tilde J'}_3 \right)} \right]   \right.  \nonumber  \\
&~~~~~~~~~~~~~~~~~~\left. - \sinh \left[ {2\beta \left( {\tilde J'}_2 - {\tilde J'}_3 \right)} \right] - 2\sinh \left( {2\beta {{\tilde J'}_3}} \right) \right\},  \nonumber  \\
&\tilde \omega_{15} = -2\sinh \left( \beta J^{\prime}_1 \right)\left\{ \sinh \left[ {2\beta \left( {\tilde J'}_2 + {\tilde J'}_3 \right)} \right]   \right.  \nonumber  \\
&~~~~~~~~~~~~~~~~~~\left. - \sinh \left[ {2\beta \left( {\tilde J'}_2 - {\tilde J'}_3 \right)} \right] - 2\sinh \left( {2\beta {{\tilde J'}_3}} \right) \right\},  \nonumber  \\
&\tilde \omega_{16} = -2\sinh \left( \beta J^{\prime}_1 \right)\left\{ \sinh \left[ {2\beta \left( {\tilde J'}_2 + {\tilde J'}_3 \right)} \right]   \right.  \nonumber  \\
&~~~~~~~~~~~~~~~~~~\left. - \sinh \left[ {2\beta \left( {\tilde J'}_2 - {\tilde J'}_3 \right)} \right] + 2\sinh \left( {2\beta {{\tilde J'}_3}} \right) \right\}. 
\label{eq34}
\end{align}
It can be examined that ${\tilde \omega }_{9}{\tilde \omega }_{10} = {\tilde \omega }_{13}{\tilde \omega }_{14}$ and ${\tilde \omega }_{11}{\tilde \omega }_{12} = {\tilde \omega }_{15}{\tilde \omega }_{16}$, such that the odd free-fermion condition Eq.~(\ref{eq12}) is satisfied. Therefore, the triangular lattice Ising model in the imaginary field is equivalent with an odd free-fermion model. Now the result is 
\begin{align}
&\mathop {\lim }\limits_{N \to \infty } \frac{1}{N}\ln Z = i\frac{\pi }{2} + \ln B + \ln A'_2 + \ln A'_3    \nonumber  \\
&~~~~~~~~~~~~~~~~~~~~~~~~~~~~~~~~~~ + \mathop {\lim }\limits_{{N_{{\rm{v}}}} \to \infty } \frac{1}{{{N_{{\rm{v}}}}}}\ln {Z_{{\rm{odd}}}},  \label{eq35}
\end{align}
where the exact solution of the odd free-fermion model is given by Eq. (21) of Ref.~\cite{RN129}
\begin{align}
&\mathop {\lim }\limits_{{N_{{\rm{v}}}} \to \infty } \frac{1}{{{N_{{\rm{v}}}}}}\ln Z_{\rm{odd}} = \frac{1}{{16{\pi ^2}}}\int_0^{2\pi } d\theta \int_0^{2\pi } d\phi \ln \left[ a + b\cos \theta \right.   \nonumber \\
&~~~~~~~~~~~\left. + c\cos \phi + d\cos \left( {\theta  - \phi } \right) + e\cos \left( {\theta  + \phi }\right) \right]  \label{eq36}
\end{align}
with 
\begin{eqnarray}
a &=& 2 \left[ \left( {\tilde \omega_9}{\tilde \omega }_{10} + {\tilde \omega }_{11}{\tilde \omega }_{12} \right)^2 + \left( {\tilde \omega }_9{\tilde \omega }_{12} \right)^2 + \left( {\tilde \omega }_{10}{\tilde \omega }_{11} \right)^2 \right. \nonumber \\
&& ~~~\left. + \left( {\tilde \omega }_{13}{\tilde \omega }_{16} \right)^2 + \left( {\tilde \omega }_{14}{\tilde \omega }_{15} \right)^2 \right],  \nonumber \\
b &=& 2\left[ -\left({\tilde \omega_9}{\tilde \omega }_{12} \right)^2 -\left({\tilde \omega_{10}}{\tilde \omega }_{11} \right)^2 + 2\tilde \omega_{13}\tilde \omega_{14}\tilde \omega_{15}\tilde \omega_{16} \right],  \nonumber \\
c &=& 2\left[ \left( \tilde \omega_{13}\tilde \omega_{16} \right)^2 + \left( \tilde \omega_{14}\tilde \omega_{15} \right)^2 - 2\tilde \omega_{9}\tilde \omega_{10}\tilde \omega_{11}\tilde \omega_{12} \right], \nonumber \\
d &=& -2 \left( \tilde \omega_9\tilde \omega_{10} - \tilde \omega_{15}\tilde \omega_{16} \right)^2, \nonumber \\
e &=& -2 \left( \tilde \omega_{11}\tilde \omega_{12} - \tilde \omega_{15}\tilde \omega_{16} \right)^2. \label{eq37}
\end{eqnarray}
We should remark that the relation between our notation of weights and that of Ref.~\cite{RN129} is 
\begin{eqnarray*}
\tilde \omega_9 = u_1,~ \tilde \omega_{10} = u_2,~ \tilde \omega_{11} = u_4,~ \tilde \omega_{12} = u_3,  \nonumber \\
\tilde \omega_{13} = u_7,~ \tilde \omega_{14} = u_8,~ \tilde \omega_{15} = u_6,~ \tilde \omega_{16} = u_5.
\end{eqnarray*}
Substituting $A'_i = \frac{1}{2}{e^{\beta J'_i}}$ and Eqs.~(\ref{eq6}), (\ref{eq7}) and (\ref{eq34}) into Eq.~(\ref{eq35}), we have the final expression 
\begin{align}
&\mathop {\lim }\limits_{N \to \infty } \frac{1}{N}\ln Z = i\frac{\pi }{2} + \frac{1}{{16{\pi ^2}}}\int_0^{2\pi } d\theta \int_0^{2\pi } d\phi  \nonumber \\
&~\ln \left[ e^{ - 4\beta \left( J_1 + J_2 + J_3 \right)} + e^{4\beta \left( J_1 + J_2 - J_3 \right)} + e^{4\beta \left( J_1 - J_2 + J_3 \right)} \right.  \nonumber \\
&~~~~~~+ e^{4\beta \left( - J_1 + J_2 + J_3 \right)} - 4  - 8\sinh^2 \left( {2\beta J_3} \right) \cos \theta   \nonumber \\
&~~~~~~\left. + 8\sinh^2 \left( {2\beta J_2} \right) \cos \phi  - 8\sinh^2 \left( {2\beta J_1} \right) \cos \left( \theta - \phi \right) \right].  \label{eq38}
\end{align}
Then we have rederived the exact solution of this case \cite{RN67, RN274}, and indicated that the model in this case can be seen as an odd free-fermion model. 

It is easy to find that this frustrated system (when $J_1=J_2=J_3>0$) in the imaginary field has a finite residual entropy in the zero temperature limit 
\begin{eqnarray}
S &=& \frac{1}{{16{\pi ^2}}}\int_0^{2\pi } d\theta \int_0^{2\pi } d\phi \ln \left\{ 3 - 2\left[ \cos \theta  - \cos \phi  \right.  \right.  \nonumber \\
&&~~~~~~~~~~~~~~~~~~~~~~~~~~~~~~~~~~~~~~~\left.  \left. + \cos \left( {\theta  - \phi } \right) \right] \right\}  \nonumber \\
&=& 0.161533~.  \label{eq39}
\end{eqnarray}
Here we set the Boltzmann constant $k_B=1$. Interestingly, the residual entropy in the imaginary field is half that in the zero field \cite{RN81} (we can also obtain the residual entropy in the zero field by Eq.~(\ref{eq30}), and compare the value with Eq.~(\ref{eq39})). 

\subsection{The Kagomé lattice}
\begin{figure} 
 \includegraphics{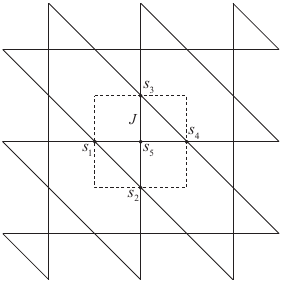}
 \caption{The diagrammatic representation of the Kagomé lattice. The region bounded by dash lines forms a vertex site of the sixteen-vertex model.} \label{fig6}
\end{figure}

Fig.~\ref{fig6} shows the Kagomé lattice. The region bounded by dash lines, which consists of two triangles, is chosen as the vertex site on the square lattice. The states of four spins $\left( s_1,s_2,s_3,s_4 \right)$ around each site can be mapped to the arrow configurations around this vertex by Eq.~(\ref{eq13}), thus the sixteen-vertex model is conveniently constructed. 

\subsubsection{Zero-field case}
The partition function is in the form of 
\begin{equation}
Z = \sum\limits_{\left\{ {\left( {{s_1},{s_2},{s_3},{s_4}} \right)} \right\} =  \pm 1}~{\prod\limits_{{\rm{all~vertices}}} {\omega \left( s_1,s_2,s_3,s_4 \right)} }   \label{eq40}
\end{equation}
with
\begin{align}
&\omega \left( {{s_1},{s_2},{s_3},{s_4}} \right) = \sum\limits_{{s_5} =  \pm 1} {e^{ - \beta {J_3}\left( {{s_1}{s_2} + {s_3}{s_4}} \right)}} \times  \nonumber  \\
&~~~~~~~~~~~~~~~~~~~~~~~~~~e^{ - \beta \left[ {{J_1}{s_5}\left( {{s_2} + {s_3}} \right) + {J_2}{s_5}\left( {{s_1} + {s_4}} \right)} \right]} .   \label{eq41}
\end{align}
The translated vertex weights from the mapping in Eq.~(\ref{eq13}) are 
\begin{align}
&{\omega _1} = {\omega _2} = {e^{ -2\beta {J_3}}}\left[ {{e^{2\beta \left( {{J_1} + {J_2}} \right)}} + {e^{ - 2\beta \left( {{J_1} + {J_2}} \right)}}} \right],  \nonumber \\
&{\omega _3} = {\omega _4} = {e^{ 2\beta {J_3}}}\left[ {{e^{2\beta \left( {{J_1} - {J_2}} \right)}} + {e^{ - 2\beta \left( {{J_1} - {J_2}} \right)}}} \right],  \nonumber \\
&{\omega _5} = {\omega _6} = 2{e^{ 2\beta J_3}},  \nonumber \\
&{\omega _7} = {\omega _8} = 2{e^{ -2\beta J_3}},  \nonumber \\
&{\omega _9} = \omega _{10} = \omega _{11} = \omega _{12} = {e^{ 2\beta J_2}} + {e^{ -2\beta J_2}},  \nonumber \\
&\omega _{13} = \omega _{14} = \omega _{15} = \omega _{16} = {e^{ 2\beta J_1}} + {e^{ -2\beta J_1}}. 
\label{eq42}
\end{align}
Again, this symmetric sixteen-vertex model are transformed into an even eight-vertex model by Eq.~(\ref{eq8}) with the weights
\begin{align}
&{\tilde \omega }_1 = {e^{ - 2\beta {J_3}}}\cosh \left[ {2\beta \left( {{J_1} + {J_2}} \right)} \right] + {e^{2\beta {J_3}}}\cosh \left[ {2\beta \left( {{J_1} - {J_2}} \right)} \right]  \nonumber  \\
&~~~~~~~+ 2\cosh \left( {2\beta {J_3}} \right) + 2\cosh \left( {2\beta {J_2}} \right) + 2\cosh \left( {2\beta {J_1}} \right),  \nonumber  \\
&{\tilde \omega }_2 = {e^{ - 2\beta {J_3}}}\cosh \left[ {2\beta \left( {{J_1} + {J_2}} \right)} \right] + {e^{2\beta {J_3}}}\cosh \left[ {2\beta \left( {{J_1} - {J_2}} \right)} \right]  \nonumber  \\
&~~~~~~~+ 2\cosh \left( {2\beta {J_3}} \right) - 2\cosh \left( {2\beta {J_2}} \right) - 2\cosh \left( {2\beta {J_1}} \right),  \nonumber  \\
&{\tilde \omega }_3 = {e^{ - 2\beta {J_3}}}\cosh \left[ {2\beta \left( {{J_1} + {J_2}} \right)} \right] + {e^{2\beta {J_3}}}\cosh \left[ {2\beta \left( {{J_1} - {J_2}} \right)} \right]  \nonumber  \\
&~~~~~~~- 2\cosh \left( {2\beta {J_3}} \right) - 2\cosh \left( {2\beta {J_2}} \right) + 2\cosh \left( {2\beta {J_1}} \right),  \nonumber  \\
&{\tilde \omega }_4 = {e^{ - 2\beta {J_3}}}\cosh \left[ {2\beta \left( {{J_1} + {J_2}} \right)} \right] + {e^{2\beta {J_3}}}\cosh \left[ {2\beta \left( {{J_1} - {J_2}} \right)} \right]  \nonumber  \\
&~~~~~~~- 2\cosh \left( {2\beta {J_3}} \right) + 2\cosh \left( {2\beta {J_2}} \right) - 2\cosh \left( {2\beta {J_1}} \right),  \nonumber  \\
&{\tilde \omega }_5 = {\tilde \omega }_6 = {e^{ - 2\beta {J_3}}}\cosh \left[ {2\beta \left( {{J_1} + {J_2}} \right)} \right]   \nonumber  \\
&~~~~~~~~~~~~~~- {e^{2\beta {J_3}}}\cosh \left[ {2\beta \left( {{J_1} - {J_2}} \right)} \right] + 2\sinh \left( {2\beta {J_3}} \right),  \nonumber  \\
&{\tilde \omega }_7 = {\tilde \omega }_8 = {e^{ - 2\beta {J_3}}}\cosh \left[ {2\beta \left( {{J_1} + {J_2}} \right)} \right]   \nonumber  \\
&~~~~~~~~~~~~~~- {e^{2\beta {J_3}}}\cosh \left[ {2\beta \left( {{J_1} - {J_2}} \right)} \right] - 2\sinh \left( {2\beta {J_3}} \right). 
\label{eq43}
\end{align}
One can examine that ${\tilde \omega }_1{\tilde \omega }_2 = {\tilde \omega }_7{\tilde \omega }_8$ and ${\tilde \omega }_3{\tilde \omega }_4 = {\tilde \omega }_5{\tilde \omega }_6$, therefore this case is an even free-fermion model. Making use of the even free-fermion solution as expressed in Eqs.~(\ref{eq18}) and (\ref{eq19}), we give the exact result
\begin{eqnarray}
&&\mathop {\lim }\limits_{N \to \infty } \frac{1}{N}\ln Z  \nonumber \\
&=&\frac{1}{3}\mathop {\lim }\limits_{N_{\rm{v}} \to \infty } \frac{1}{{{N_{{\rm{v}}}}}}\ln Z_{\rm{even}}  \nonumber \\
&=&\frac{1}{24{\pi^2}}\int_0^{2\pi } d\theta \int_0^{2\pi } d\phi \ln\left\{ 16\left[ \left( C_1C_2C_3 - S_1S_2S_3 \right)^2  \right.  \right.   \nonumber \\
&& \left. + C_1^2 + C_2^2 + C_3^2 \right] +32S_1\left(S_1C_2C_3 - C_1S_2S_3 \right)\cos \theta   \nonumber \\
&&~~+32S_2\left(C_1S_2C_3 - S_1C_2S_3 \right)\cos \phi   \nonumber \\
&&~~\left. -32S_3\left(C_1C_2S_3 - S_1S_2C_3 \right) \cos \left( \theta - \phi \right) \right\}  \label{eq44}
\end{eqnarray}
with $C_i=\cosh(2\beta J_i)$ and $S_i=\sinh(2\beta J_i)$ ($i=1,2,3$). This result agrees with the known expressions (see Eq. (36) of Ref.~\cite{RN82} for the isotropic case for instance).   

\subsubsection{Imaginary-field case}
Taking into account Eq.~(\ref{eq22}), the partition function in the presence of the imaginary field is
\begin{equation}
Z = i^N\sum\limits_{\left\{ {\left( {{s_1},{s_2},{s_3},{s_4}} \right)} \right\} =  \pm 1}~{\prod\limits_{{\rm{all~vertices}}} {\omega \left( s_1,s_2,s_3,s_4 \right)} }   \label{eq45}
\end{equation}
with
\begin{eqnarray}
&\omega \left( {{s_1},{s_2},{s_3},{s_4}} \right) = \sum\limits_{s_5 = \pm 1}(s_1s_2s_5) e^{ - \beta {J_3}\left( {{s_1}{s_2} + {s_3}{s_4}} \right)} \times  \nonumber  \\
&~~~~~~~~~~~~~~~~~~~~~~~~~~e^{ - \beta \left[ {{J_1}{s_5}\left( {{s_2} + {s_3}} \right) + {J_2}{s_5}\left( {{s_1} + {s_4}} \right)} \right]}.   \label{eq46}
\end{eqnarray}
We find that the resulting sixteen-vertex model is antisymmetric, like in the case of triangular lattice model in the imaginary field. The vertex weights are listed
\begin{align}
& {\omega _1} = {e^{ -2\beta {J_3}}}\left[ e^{ - 2\beta \left( J_1 + J_2 \right)} - e^{2\beta \left( J_1 + J_2 \right)} \right],  \nonumber \\
& \omega _2 = -\omega _1,  \nonumber \\
& {\omega _3} = {e^{ 2\beta {J_3}}}\left[ e^{ - 2\beta \left( J_1 - J_2 \right)} - e^{2\beta \left( J_1 - J_2 \right)} \right],  \nonumber \\
& \omega _4 = -\omega _3,  \nonumber \\
& {\omega _5} = {\omega _6} = {\omega _7} = {\omega _8} = 0,  \nonumber \\
& \omega _9 =\omega _{11} = e^{ 2\beta J_2} - e^{ -2\beta J_2},  \nonumber \\
& \omega _{10} = \omega _{12} = -\omega _9,  \nonumber \\
& \omega _{13} =\omega _{15} = e^{ 2\beta J_1} - e^{ -2\beta J_1},  \nonumber \\
& \omega _{14} = \omega _{16} = -\omega _{13}. 
\label{eq47}
\end{align}
Then the transformed weights of the odd eight-vertex model are 
\begin{align}
& \tilde \omega_9 = {e^{ - 2\beta {J_3}}}\sinh \left[ {2\beta \left( {{J_1} + {J_2}} \right)} \right] - {e^{2\beta {J_3}}}\sinh \left[ {2\beta \left( {{J_1} - {J_2}} \right)} \right]  \nonumber  \\
&~~~~~~~ +2\sinh(2\beta J_2),  \nonumber  \\
& \tilde \omega_{10} = {e^{ - 2\beta {J_3}}}\sinh \left[ {2\beta \left( {{J_1} + {J_2}} \right)} \right] - {e^{2\beta {J_3}}}\sinh \left[ {2\beta \left( {{J_1} - {J_2}} \right)} \right]  \nonumber  \\
&~~~~~~~~ -2\sinh(2\beta J_2),  \nonumber  \\
& \tilde \omega_{11} = \tilde \omega_{9},   \nonumber  \\
& \tilde \omega_{12} = \tilde \omega_{10},  \nonumber  \\
& \tilde \omega_{13} = {e^{ - 2\beta {J_3}}}\sinh \left[ {2\beta \left( {{J_1} + {J_2}} \right)} \right] + {e^{2\beta {J_3}}}\sinh \left[ {2\beta \left( {{J_1} - {J_2}} \right)} \right]  \nonumber  \\
&~~~~~~~~ +2\sinh(2\beta J_1),  \nonumber  \\
& \tilde \omega_{14} = {e^{ - 2\beta {J_3}}}\sinh \left[ {2\beta \left( {{J_1} + {J_2}} \right)} \right] + {e^{2\beta {J_3}}}\sinh \left[ {2\beta \left( {{J_1} - {J_2}} \right)} \right]  \nonumber  \\
&~~~~~~~~ -2\sinh(2\beta J_1),  \nonumber  \\
& \tilde \omega_{15} = \tilde \omega_{13},  \nonumber  \\
& \tilde \omega_{16} = \tilde \omega_{14}. 
\label{eq48}
\end{align}
It is straightforward to verify that ${\tilde \omega }_{9}{\tilde \omega }_{10} = {\tilde \omega }_{13}{\tilde \omega }_{14}$ such that this case is equivalent with an odd free-fermion model, the exact result is thereby given by the odd free-fermion solution in Eqs.~(\ref{eq36}) and (\ref{eq37})
\begin{align}
&~~~ \mathop {\lim }\limits_{N \to \infty } \frac{1}{N}\ln Z  \nonumber \\
& =  i\frac{\pi }{2} + \frac{1}{3}\mathop {\lim }\limits_{N_{\rm{v}} \to \infty } \frac{1}{N_{\rm{v}}}\ln Z_{\rm{odd}}  \nonumber \\
& =  i\frac{\pi }{2} + \frac{1}{6}\ln \left[ 16 \left| 2 S_1^2 S_2^2 S_3^2 + S_1^2 S_2^2 + S_1^2 S_3^2 + S_2^2 S_3^2 \right. \right.  \nonumber \\
&~~~~~~~~~~~~~~~~~~~~~~ \left. \left. - 2 S_1S_2S_3C_1C_2C_3 \right| \right]  \label{eq49}
\end{align}
with $C_i=\cosh(2\beta J_i)$ and $S_i=\sinh(2\beta J_i)$ ($i=1,2,3$). The expression is surprisingly simple. To our knowlegde, this result has not been reported previously. There have been research works studying the Kagomé lattice Ising model in a magnetic field by transformation into the honeycomb lattice model \cite{RN266, RN302, RN296}. One can examine that, the case in the imaginary field was not included in these works. Our analysis demonstrates that this case is also exactly solvable.

\blue{We note that, the details of the mapping and calculation procedure are very important. In this case, the vertex weights are defined in Eq. (\ref{eq46}), in which the factor $\prod\limits_{i = 1}^N {{s_i}}$ in the partition function exhibits a subfactor ${s_1}{s_2}{s_5}$. One can see that, the effect of the field is partitioned as ${e^{\beta {H_{{\rm{ex}}}}\left( {{s_1} + {s_2} + {s_5}} \right)}}$ for this vertex unit. Now we repartition this effect as ${e^{\beta {H_{{\rm{ex}}}}\left[ {\frac{1}{2}\left( {{s_1} + {s_2} + {s_3} + {s_4}} \right) + {s_5}} \right]}}$, i.e., the vertex weights are reset as
\begin{align*}
&\omega \left( {{s_1},{s_2},{s_3},{s_4}} \right) = \sum\limits_{{s_5} =  \pm 1} \left( s_1^{1/2}s_2^{1/2}s_3^{1/2}s_4^{1/2}{s_5} \right) \times  \nonumber \\
&~~~~~~~~~~~~~~~~{e^{ - \beta {J_3}\left( {{s_1}{s_2} + {s_3}{s_4}} \right)}}{e^{ - \beta \left[ {{J_1}{s_5}\left( {{s_2} + {s_3}} \right) + {J_2}{s_5}\left( {{s_1} + {s_4}} \right)} \right]}} 
\end{align*} 
with $1^{1/2} = 1$ and ${(-1)}^{1/2} = i$. In this way, we will find that
\begin{align*}
&{\omega _1} =  - {\omega _2},{\rm{ }}{\omega _3} =  - {\omega _4},{\rm{ }}{\omega _5} = {\omega _6} = {\omega _7} = {\omega _8} = 0, \nonumber \\
&{\omega _9} = {\omega _{10}},{\rm{ }}{\omega _{11}} = {\omega _{12}},{\rm{ }}{\omega _{13}} = {\omega _{14}},{\rm{ }}{\omega _{15}} = {\omega _{16}}.
\end{align*}
We obtain neither a symmetric nor an antisymmetric sixteen-vertex model. Hence, we fail to map the Ising model into an even or odd free-fermion model by weak-graph expansion. We will probably miss this new solvable case, or at least are not able to solve this case so conveniently by mapping into the odd free-fermion model as we have shown. This fact indicates that, the details of mapping and calculation may play the key role in leading to the new solution.
}

As is well-known, most of the exact solutions of two-dimensional Ising models are of the form of a double integral. Onsager's solution of the square lattice model \cite{RN72} is a very early example. Each model studied in this paper using the free-fermion approach involves a double integral in the solution, except for the Kagomé lattice model in the imaginary field. It is interesting to compare the result Eq.~(\ref{eq49}) with Eq. (1) of Ref.~\cite{RN144}, which is the solution of the dimer model on the Kagomé lattice and can also be derived by mapping to the odd free-fermion model. Both formulas take a very simple logarithmic form. Ref.~\cite{RN144} stated that the novel and unique expression ``points to the special role played by the Kagomé lattice''. Here we provide new evidence for this statement.

When $J_1=J_2=J_3>0$, the Kagomé lattice Ising model is geometrically frustrated. The residual entropy in the imaginary field is easily calculated from Eq.~(\ref{eq49}) by taking the zero temperature limit
\begin{equation}
S =\frac{1}{6}\ln 3 = 0.183102.   \label{eq50}
\end{equation}

\section{Summary}   \label{summary}
The exact solutions of Ising models on typical two-dimensional lattices, specifically the honeycomb, triangular and Kagomé lattices, are studied in the free-fermion formulation. For each Ising model, both the case of the zero field and the case of an imaginary field $i( {{\pi  \mathord{\left/{\vphantom {\pi  2}} \right. \kern-\nulldelimiterspace} 2}} )k_BT$ are considered. Five known solutions are rederived and one new solution is found. In particular, the Kagomé lattice model in the imaginary field is shown to be exactly solvable, exhibiting a very simple form of solution. Hence, we have introduced a new member to the family of exactly solvable lattice models.

All solutions in this paper are achieved by mapping the Ising spin system into a free-fermion model. The mapping procedure is straightforward and should be easy to extend to other lattice systems. Applying the mapping method to the Ising model in a physical (real) and non-zero magnetic field would be interesting. Furthermore, the free-fermion formulation may be suitable for other statistical lattice problems, e.g., the dimer covering problem. Our work has provided new insights into the mathematically exact results of spin systems, showcasing the elegance and power of the free-fermion formulation in condensed matter and statistical physics.

\begin{acknowledgments}
This work was supported by Guangdong Provincial Quantum Science Strategic Initiative (Grants No. GDZX2203001 and No. GDZX2403001), National Natural Science Foundation of China (Grants No. 12474489 and No. 12474228), Research Funding for Outbound Postdoctoral Fellows in Shenzhen (Grant No. SZRCXM2401006), Shenzhen Fundamental Research Program (Grant No. JCYJ20240813153139050), and Innovation Program for Quantum Science and Technology (Grant No. 2021ZD0302300).
\end{acknowledgments}

\bibliographystyle{apsrev4-1}

\end{document}